\documentstyle[12pt,psfig,epsf]{article}
\def\today{June 20, 1996}
\newcommand{\getepl}[1]{\centerline{\epsfxsize=4in\epsfbox{#1}\vspace{-37pt}}}

\def\FIGA{%
\begin{figure}
	\getepl{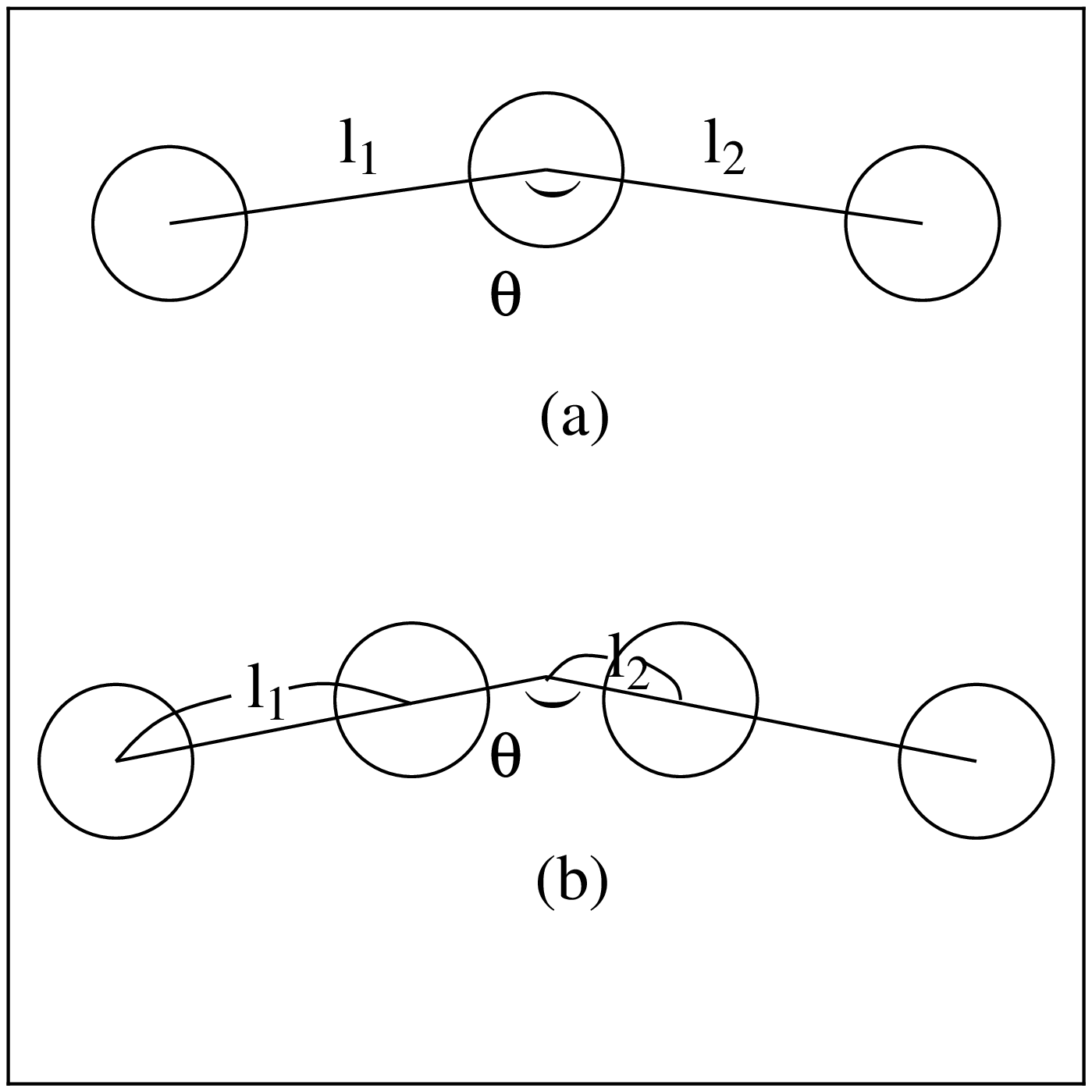}
	\caption{The linear-chain configurations of 3$\alpha$ (a) 
         and 4$\alpha$ (b).}
	\label{lincon}
\end{figure}}
\def\FIGB{%
\begin{figure}
	\getepl{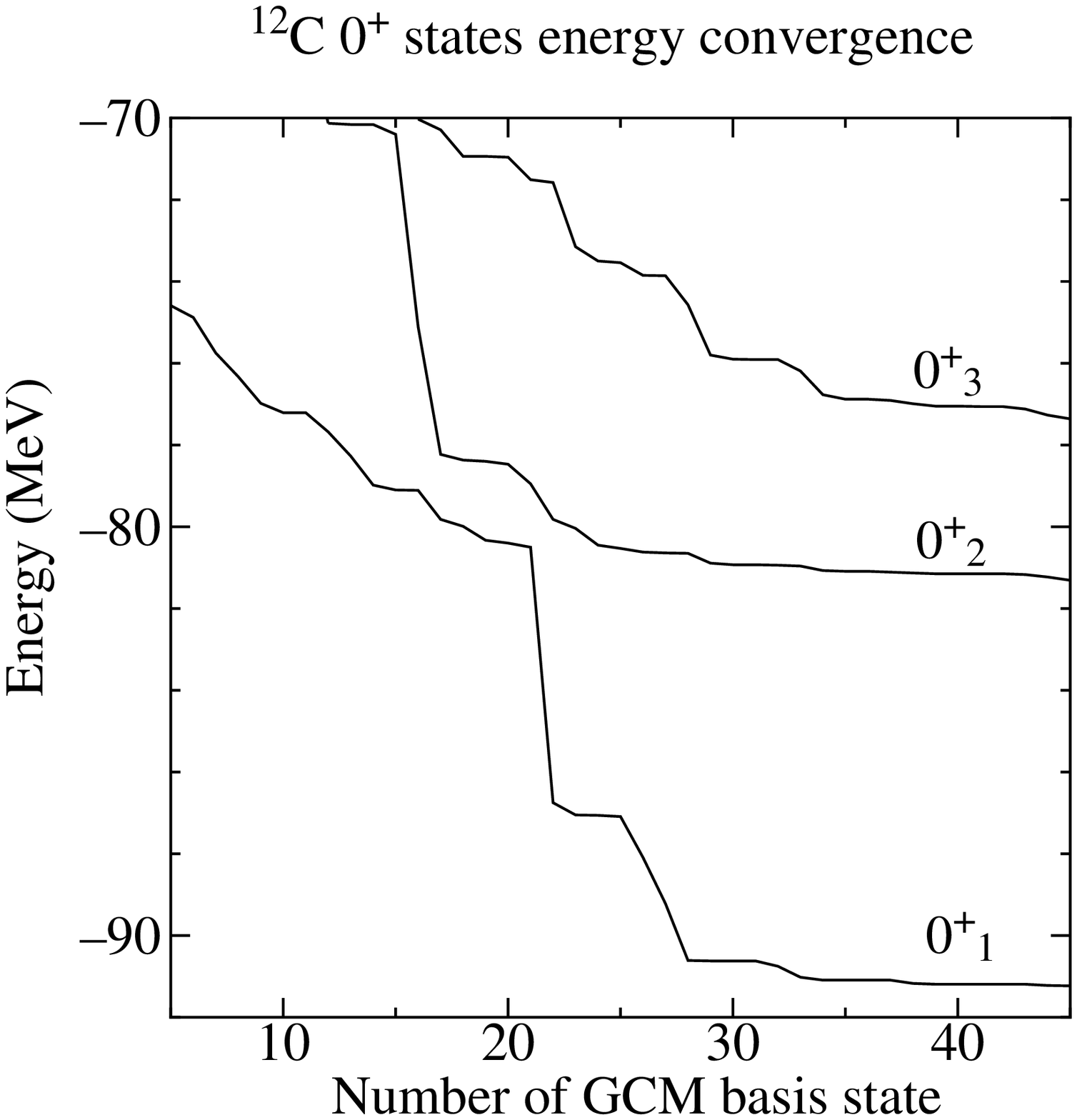}
	\caption{The diagonalization of Hamiltonian of \nuc{12}{C}.}
	\label{Cdiago}
\end{figure}}
\def\FIGC{%
\begin{figure}
	\getepl{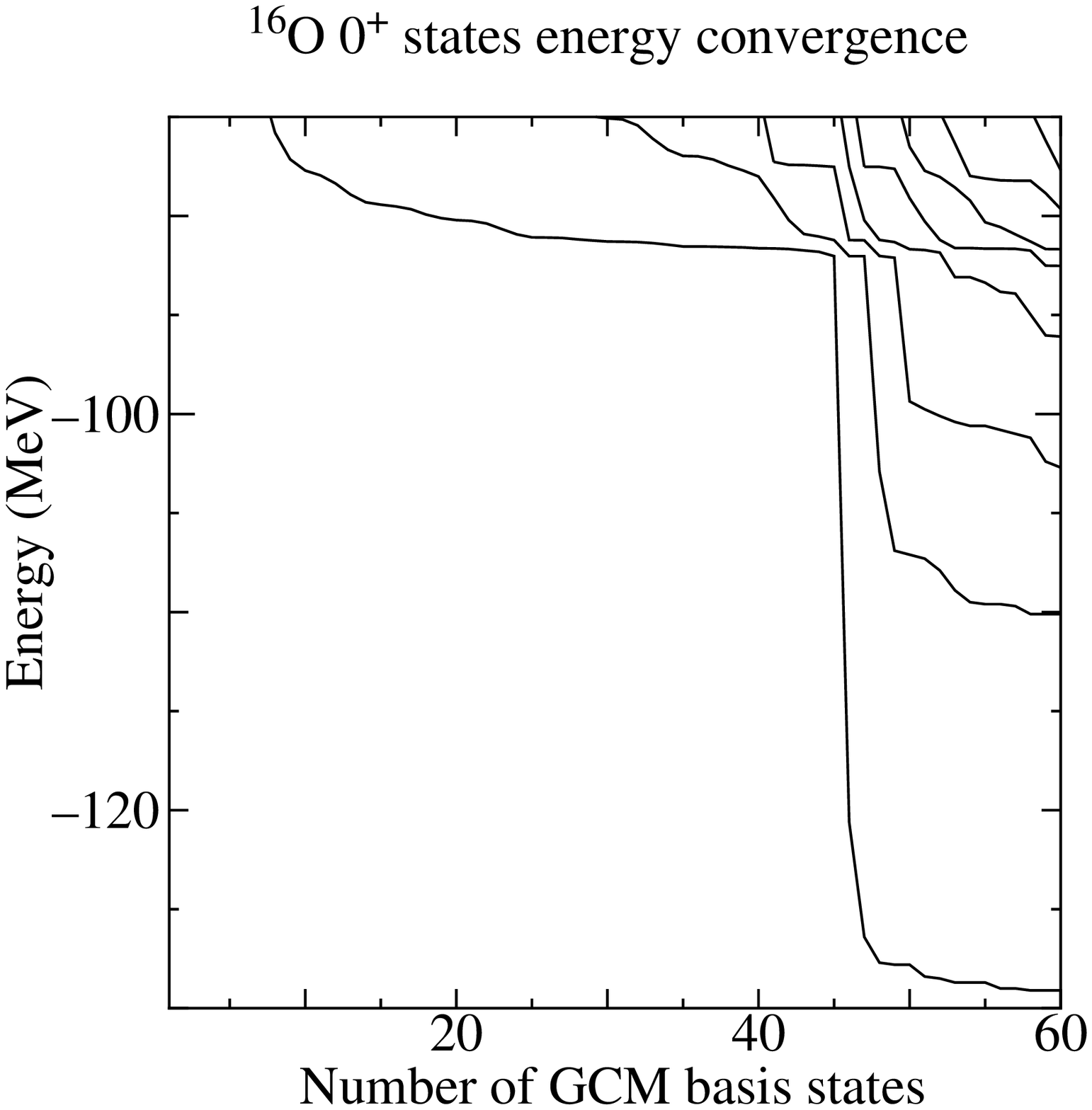}
	\caption{The diagonalization of Hamiltonian of \nuc{16}{O}.}
	\label{Odiago}
\end{figure}}
\def\FIGD{%
\begin{figure}
	\getepl{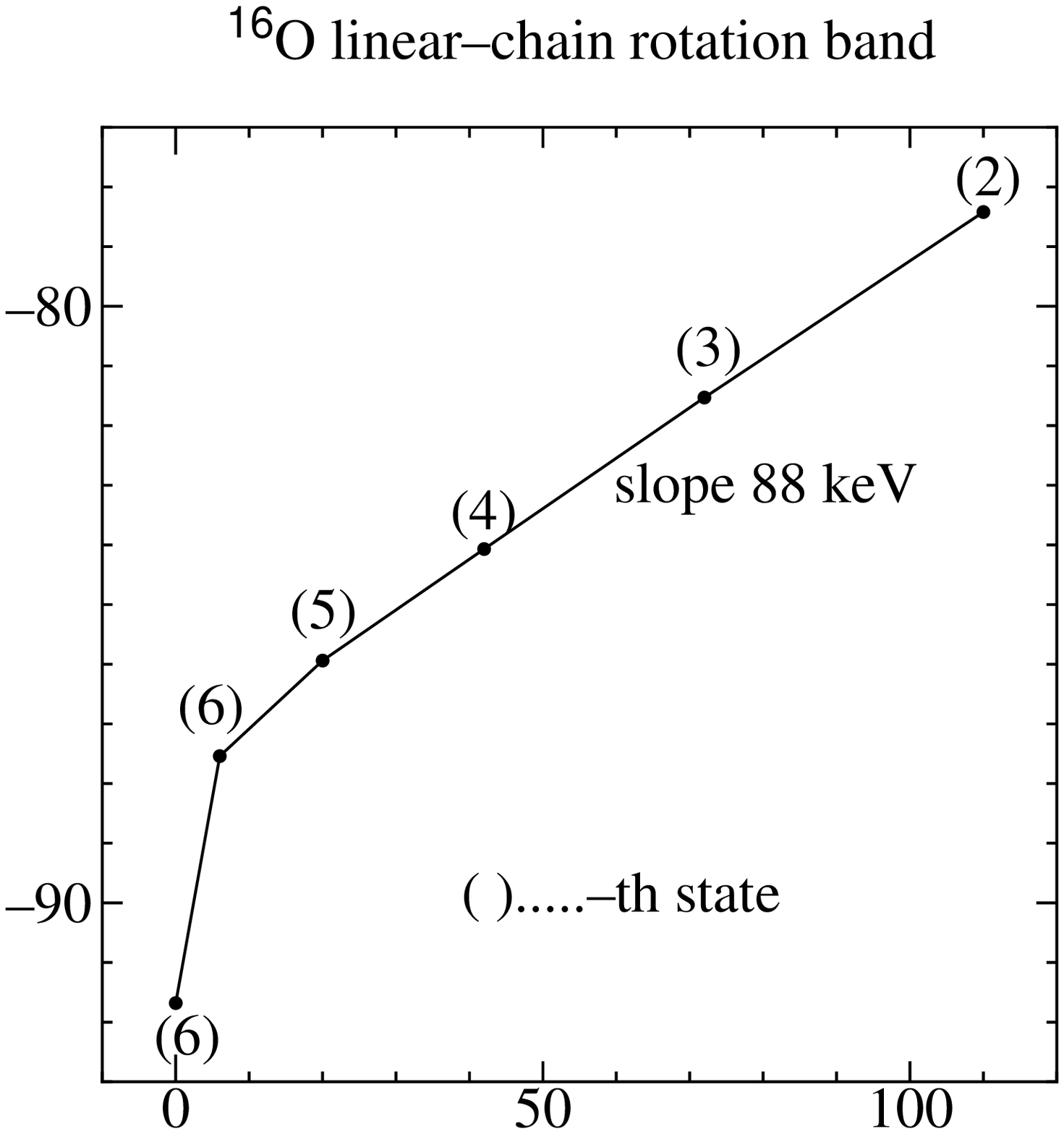}
	\caption{The rotational band of \nuc{16}{O} linear-chain state.
        The numbers show the order of excited state.}
	\label{Orot}
\end{figure}}

\def\TABLEA{%
\begin{table}
\centerline{
\begin{tabular}{r|c|c|c||r|c|c|c||r|c|c|c}
\hline
\hline
 & $l_1$ & $l_2$ & $\theta$ &  
 & $l_1$ & $l_2$ & $\theta$ & 
 & $l_1$ & $l_2$ & $\theta$ \\
\hline 
\hline 
 1 & 1.0 & 3.5 & $\pi$      & 16 & 1.0 & 4.7 & $9\pi/12$ & 31 & 1.0 & 1.4 & $6\pi/12$\\ 
 2 & 2.0 & 3.2 & $\pi$      & 17 & 2.0 & 2.9 & $9\pi/12$ & 32 & 2.0 & 1.9 & $6\pi/12$\\ 
 3 & 3.0 & 3.0 & $\pi$      & 18 & 3.0 & 2.9 & $9\pi/12$ & 33 & 3.0 & 2.4 & $6\pi/12$\\ 
 4 & 4.0 & 2.9 & $\pi$      & 19 & 4.0 & 2.9 & $9\pi/12$ & 34 & 4.0 & 2.7 & $6\pi/12$\\ 
 5 & 5.0 & 2.9 & $\pi$      & 20 & 5.0 & 2.9 & $9\pi/12$ & 35 & 5.0 & 2.9 & $6\pi/12$\\ 
 6 & 1.0 & 3.5 & $11\pi/12$ & 21 & 1.0 & 4.5 & $8\pi/12$ & 36 & 1.0 & 1.5 & $5\pi/12$\\ 
 7 & 2.0 & 3.2 & $11\pi/12$ & 22 & 2.0 & 2.5 & $8\pi/12$ & 37 & 2.0 & 2.0 & $5\pi/12$\\ 
 8 & 3.0 & 3.0 & $11\pi/12$ & 23 & 3.0 & 2.7 & $8\pi/12$ & 38 & 3.0 & 2.4 & $5\pi/12$\\ 
 9 & 4.0 & 2.9 & $11\pi/12$ & 24 & 4.0 & 2.9 & $8\pi/12$ & 39 & 4.0 & 2.7 & $5\pi/12$\\ 
10 & 5.0 & 2.9 & $11\pi/12$ & 25 & 5.0 & 2.9 & $8\pi/12$ & 40 & 5.0 & 2.9 & $5\pi/12$\\ 
11 & 1.0 & 3.3 & $10\pi/12$ & 26 & 1.0 & 4.9 & $7\pi/12$ & 41 & 1.0 & 1.3 & $4\pi/12$\\ 
12 & 2.0 & 3.1 & $10\pi/12$ & 27 & 2.0 & 2.2 & $7\pi/12$ & 42 & 2.0 & 2.1 & $4\pi/12$\\ 
13 & 3.0 & 2.9 & $10\pi/12$ & 28 & 3.0 & 2.5 & $7\pi/12$ & 43 & 3.0 & 2.6 & $4\pi/12$\\ 
14 & 4.0 & 2.9 & $10\pi/12$ & 29 & 4.0 & 2.8 & $7\pi/12$ & 44 & 4.0 & 2.8 & $4\pi/12$\\
15 & 5.0 & 2.9 & $10\pi/12$ & 30 & 5.0 & 2.9 & $7\pi/12$ & 45 & 5.0 & 2.9 & $4\pi/12$\\ 
\hline  
\end{tabular}
}
	\caption{ The GCM basis states for \nuc{12}{C} described by 
	$l_1$ (fm), $l_2$ (fm) and $\theta$ (radian) of Fig. 1 (a). 
	$l_2$ is determined to minimize the energy at given $l_1$ and $\theta$.}
\end{table}}

\def\TABLEB{
\begin{table}
\centerline{
\begin{tabular}{r|c|c|c||r|c|c|c||r|c|c|c||r|c|c|c}
\hline
\hline
 & $l_1$ & $l_2$ & $\theta$ &  
 & $l_1$ & $l_2$ & $\theta$ & 
 & $l_1$ & $l_2$ & $\theta$ & 
 & $n_x$ & $n_y$ & $n_z$ \\
\hline 
\hline 
 1 & 0.5 & 3.8 & $\pi$      & 16 & 0.5 & 3.8 & $9\pi/12$ & 31 & 0.5 & 3.4 & $6\pi/12$ & 46 & 4 & 4 & 5 \\ 
 2 & 1.0 & 3.5 & $\pi$      & 17 & 1.0 & 3.5 & $9\pi/12$ & 32 & 1.0 & 3.3 & $6\pi/12$ & 47 & 4 & 4 & 6 \\ 
 3 & 1.5 & 3.3 & $\pi$      & 18 & 1.5 & 3.3 & $9\pi/12$ & 33 & 1.5 & 3.3 & $6\pi/12$ & 48 & 4 & 4 & 8 \\ 
 4 & 2.0 & 3.2 & $\pi$      & 19 & 2.0 & 3.2 & $9\pi/12$ & 34 & 2.0 & 3.2 & $6\pi/12$ & 49 & 4 & 4 &10 \\ 
 5 & 2.5 & 3.2 & $\pi$      & 20 & 2.5 & 3.2 & $9\pi/12$ & 35 & 2.5 & 3.2 & $6\pi/12$ & 50 & 4 & 4 &15 \\ 
 6 & 0.5 & 3.8 & $11\pi/12$ & 21 & 0.5 & 3.7 & $8\pi/12$ & 36 & 0.5 & 3.0 & $5\pi/12$ & 51 & 5 & 5 & 5 \\ 
 7 & 1.0 & 3.5 & $11\pi/12$ & 22 & 1.0 & 3.5 & $8\pi/12$ & 37 & 1.0 & 3.1 & $5\pi/12$ & 52 & 5 & 5 & 7 \\ 
 8 & 1.5 & 3.3 & $11\pi/12$ & 23 & 1.5 & 3.3 & $8\pi/12$ & 38 & 1.5 & 3.1 & $5\pi/12$ & 53 & 5 & 5 & 9 \\ 
 9 & 2.0 & 3.2 & $11\pi/12$ & 24 & 2.0 & 3.2 & $8\pi/12$ & 39 & 2.0 & 3.1 & $5\pi/12$ & 54 & 5 & 5 &12 \\ 
10 & 2.5 & 3.2 & $11\pi/12$ & 25 & 2.5 & 3.2 & $8\pi/12$ & 40 & 2.5 & 3.2 & $5\pi/12$ & 55 & 5 & 5 &15 \\ 
11 & 0.5 & 3.8 & $10\pi/12$ & 26 & 0.5 & 3.6 & $7\pi/12$ & 41 & 0.5 & 2.9 & $4\pi/12$ & 56 & 6 & 6 & 6 \\ 
12 & 1.0 & 3.5 & $10\pi/12$ & 27 & 1.0 & 3.4 & $7\pi/12$ & 42 & 1.0 & 2.9 & $4\pi/12$ & 57 & 6 & 6 & 8 \\ 
13 & 1.5 & 3.3 & $10\pi/12$ & 28 & 1.5 & 3.3 & $7\pi/12$ & 43 & 1.5 & 2.9 & $4\pi/12$ & 58 & 6 & 6 &10 \\ 
14 & 2.0 & 3.2 & $10\pi/12$ & 29 & 2.0 & 3.2 & $7\pi/12$ & 44 & 2.0 & 3.0 & $4\pi/12$ & 59 & 6 & 6 &15 \\ 
15 & 2.5 & 3.2 & $10\pi/12$ & 30 & 2.5 & 3.2 & $7\pi/12$ & 45 & 2.5 & 3.0 & $4\pi/12$ & 60 & 6 & 6 &20 \\ 
\hline 
\end{tabular}
}
	\caption{The GCM basis states for \nuc{16}{O}. 
	Basis state configurations configuration No.1 $\sim$ No.45 
	are described by $l_1$ (fm) $l_2$ (fm)
	and $\theta$ (radian) of Fig. 1 (b). The value of
	$l_2$ is determined to minimize the
	total energy. The GCM basis states 
	No.46 $\sim$ No.60 are obtained by the constraint cooling method, and 
	classified with the
	constraint principal quantum numbers ($n_x$, $n_y$ and $n_z$).
	}
\end{table}}

\newcommand{\TITLE}[3]{%
\begin{center}{\Large {#1}}\\ \vspace{0.5cm}{#2}\\ 
\vspace{0.3cm}{\it #3}\end{center}\vspace{0.3cm}}%

\def\ack{\section*{Acknowledgement}%
  \addtocontents{toc}{\protect\vspace{6pt}}%
  \addcontentsline{toc}{section}{Acknowledgement}}
\def\nuc#1#2{\relax\ifmmode{}^{#1}{\protect\text{#2}}\else${}^{#1}$#2\fi}
\newcommand{\del}{\partial}                     %       partial
\newcommand{\Hml}{{\cal H}}		%       Hamiltonian

\newcommand{\comment}[1]{}

\newcommand{\etal}{{\it et al.}}

\newcommand{\beqar}{\begin{eqnarray}}
\newcommand{\eeqar}{\end{eqnarray}}
\newcommand{\beq}{\begin{equation}}
\newcommand{\eeq}{\end{equation}}
\begin{document}
\begin{titlepage}
\centerline{
Submitted to {\sl Phys. Lett. B}
\hfill
HUPS-96-4
}

\vfill

\TITLE{
	Four-Alpha Linear-Chain States in \nuc{16}{O}
}{
	Naoyuki ITAGAKI\footnote{
                E-mail: itagaki@nucl.phys.hokudai.ac.jp,\,\
                Fax: +81-11-746-5444.}, 
	Akira OHNISHI and Kiyoshi KAT\=O
}{	
	Department of Physics, Hokkaido University, Sapporo 060, Japan
}

\vfill

\begin{center}
\today
\end{center}

\vfill

\begin{abstract}
We study 4$\alpha$ linear-chain states in \nuc{16}{O} in comparison with
3$\alpha$ states in \nuc{12}{C} by using
the Generator Coordinate Method
within a microscopic $N\alpha$-cluster model.
It is shown to be very important
to solve relative motion of $\alpha$-clusters
by taking into account the orthogonality between
the linear-chain states and other low-lying states
including the ground state.
For \nuc{12}{C}, the 3$\alpha$ chain state disappears
because of this orthogonality.
The $4\alpha$ chain state in \nuc{16}{O}, on the other hand, 
is hardly affected by low-lying states
and persists to remain above the 4$\alpha$ threshold.
The calculated moment of inertia of the $4\alpha$ linear-chain rotational band
(88 keV) reproduces the experimentally suggested value qualitatively.
\end{abstract}

\vfill

\end{titlepage}
%%%
%     Introduction
%
%\newpage

\section{Introduction}

%
% Large deformation = frontire in nuclear spectroscopy
%
A nuclear study of a large deformation is one of the frontier in nuclear
spectroscopy.
Since the first observation of the superdeformed band~\cite{SD1},
various superdeformed nuclei ($\sim$ 1:2 deformation)
have been found,
and nowadays extensive studies on these bands including
their excited bands are in progress.
Furthermore, there are several reports on the hyperdeformed band
in which the deformation is around 1:3~\cite{HyperDef}.
These large deformed nuclei are mainly found
in the medium-heavy mass number region
where the liquid-drop energy and the shell-correction balances.
Therefore, larger deformation than 1:3 is not expected to appear in
this mass number region,
since the shell-correction is not enough to make a barrier
at larger deformations.

%
% Large deformation in light nuclei
%
On the other hand,
a large deformation above 1:3 may be possible in light nuclei
due to the $\alpha$-cluster structure.
For example, the first example of the ``superdeformed" nuclei
is nothing but ${}^8$Be which has the $\alpha-\alpha$ structure,
and there have been a lot of discussion on the nature of $0^+_2$ state
of $^{12}$C as the $3\alpha$ linear-chain state~\cite{Morinaga56,Morinaga66}.
%
% Present status of Large Deformed States in Light Nuclei
%
Experimentally, it is more difficult %and challenging
to observe large-deformed states or linear-chain states of 
$\alpha$-clusters in light nuclei,
since these states have large particle decay widths,
and consequently a coincident detection of multi-$\alpha$ particles
becomes necessary.
From a theoretical point of view, on the other hand,
it is required to estimate the couple-channel effects
over the broad range of deformation or geometric structure,
since a mean field picture may not be valid for the highly excited
states in light nuclei.
However,
since it gives us the opportunity to explore the {\em most deformed} nuclei, 
the study of these $N\alpha$ linear-chain states~\cite{Ikeda68} is valuable.

%
% Ultra-deformation (>= 1:4) in light nuclei
%
For example,
there are several experimental candidates in light 4$N$ nuclei
for large deformed states above 1:3.
The first example is the 4$\alpha$ linear-chain band
starting around the 4$\alpha$ threshold energy region in $^{16}$O 
suggested by Chevallier \etal~\cite{Cheva} through the data analysis 
of the \nuc{12}{C}+$\alpha$ $\to$ \nuc{8}{Be}+\nuc{8}{Be} reaction.
This suggestion is supported 
by a theoretical work by Suzuki \etal~\cite{Suzuki72}; 
they analyzed the decay widths
from such a strongly coupled linear-chain state,
and discussed that the excited states observed by Chevallier \etal \  
are possibly characterized
by a 4$\alpha$ linear-chain structure.
%
% 1:6 ????
%
The second example is the 6$\alpha$ linear-chain states, 
which is discussed as a candidate to explain a new molecular resonance state
observed by Wuosmma \etal~\cite{Wuosmma}
in the excitation function of the inelastic reaction
	\nuc{12}{C}(\nuc{12}{C}, \nuc{12}{C}($0^+_2$))\nuc{12}{C}($0^+_2$)
at $E_x$=56.4 MeV~\cite{Rea92}.
Although the clarification of the existence of a 6$\alpha$ linear-chain state
is very challenging, 
the coupling to $^{12}$C$^*$--$^{12}$C$^*$ and $^{16}$O$^*$--$^8$Be
may be important~\cite{Hirabayashi}
and makes it difficult to study a pure linear-chain state.
Therefore, it is important and urgent to study the stability of
linear-chain states in \nuc{12}{C} and \nuc{16}{O} at this stage.

%
% Previous study on Linear Chains
%
The reliability of the linear-chain structure has been usually studied
through kinematical analyses of decay widths.
It has been also discussed on the analyses of small vibration 
around the equilibrium configuration by Ikeda~\cite{Horiuchi72}.
% 
%
% Is it necessary to comment the details of previous works in a letter
% especially in the introduction ?
%
%Their conclusion is that the 
%4$\alpha$ linear-chain configuration is stabilized by the repulsive 
%Coulomb interaction between $\alpha$-clusters but the 3$\alpha$ linear-chain 
%configuration cannot keep its stability.
%
In those previous studies, however,
the linear-chain structure was assumed {\it a priori}.
Namely, the coupling effects with other states have not been taken into account.
These couplings, especially with lower excited states are expected to play 
an important role on the stability of linear-chain states,
since they push-up the energy of linear-chain states
and generally increase the instability.
Therefore, it is necessary to calculate in a wide wave function space,
which covers not only the linear-chain configurations
but also the lower excited states.

%
% Purpose of this letter
%
In this paper, we investigate the stability of 
4$\alpha$ linear-chain configurations in \nuc{16}{O}
comparing with 3$\alpha$ states of \nuc{12}{C}
by applying a microscopic $N\alpha$-cluster model. % to those nuclei. <-Obvious
In contrast to the previous works, 
we discuss the stability of the $N\alpha$ linear-chain structure
% without any assumption of such a configuration but <-- ?
by solving dynamics of $N\alpha$ systems.
This is achieved by diagonalizing the Hamiltonian matrix
in a wide space which also covers
low-lying levels including the ground state.
The framework used here is based on the Generator Coordinate Method (GCM)
by which we can solve relative motion of $\alpha$ clusters
by taking account of the superposition of many different
intrinsic configurations.
We use the Constraint Cooling Method~\cite{Horiuchi92,Enyo95,Enyo95b,Enyo95c}
proposed in the framework of AMD~\cite{Ono92}
in order to generate GCM basis which describes low-lying levels.
This method has been already shown to provide the suitable GCM basis states
effectively~\cite{Ita}.

This paper is organized as follows; in section \ref{sec:Method},
we explain our model especially
the way to construct the GCM basis states.
In section \ref{sec:results},
we show our results that 
the \nuc{12}{C} linear-chain state disappears because of the orthogonality
to other states,
however the \nuc{16}{O} linear-chain state exists above 
the 4$\alpha$ threshold and shows the rotational band structure.
In section \ref{sec:summary} summary and conclusion are presented.

%\newpage
\section{Method}
\label{sec:Method}

In the framework of GCM,
the energy spectra and the wave functions are solved by 
diagonalizing the Hamiltonian matrix through the Hill-Wheeler equation;
\beq
\label{GCMwf}
\Psi^{JM}
	=	\sum_{j} c_j \Phi^{JM}_j \ ,
\eeq
\beq
\label{GCM}
\sum_{k}(
	\langle {\Phi^{JM}_j}|H|{\Phi^{JM}_k}\rangle
-E\langle{\Phi^{JM}_j}|{\Phi^{JM}_k}\rangle)c_k =0\ ,
\eeq
where $\Phi^{JM}_j$ is the $j$th GCM basis wave functions 
with the total angular momentum $J$ and its $z$-component $M$.
Since the reliability of the GCM calculation largely depends
on the adopted basis wave functions,
we have adopted the GCM basis states which reasonably reproduce
the low-lying excited states
in addition to the linear-chain configurations described in the next section.
The first set of GCM basis states is generated 
by applying the constraint cooling method~\cite{Enyo95}
to the Bloch-Brink wave function for an $N\alpha$ system.
The angular momentum projection from these wave packets is carried out
numerically.
A detailed explanation of this method is found in our previous paper~\cite{Ita}.

Hereafter, we briefly summarize our method.
%
% Wave Function
%
The Bloch-Brink wave function for an $N\alpha$ system
is the Slater determinant of Gaussian wave packets~\cite{Horiuchi92,Ono92}
with the assumption of the [4]-symmetry for the spin and iso-spin part,
\beq
\label{BB}
\Phi (\{ {\bf r}_i^{\alpha}; {\bf Z}_i \}) 
	=	{\cal A} \prod_{i=1}^N \prod_{\alpha=1}^4 
		\phi_i^{\alpha}({\bf r}_i^{\alpha}; {\bf Z}_i)\ ,
\eeq
\beq
\phi_i^{\alpha}({\bf r}^{\alpha};  {\bf Z}_i)
	= {\exp\{-\nu({\bf r}^{\alpha}-{\bf Z}_i/\sqrt{\nu})^2+{\bf Z}_i^2/2\}}
		\chi_\alpha\, \quad (i=1 \sim N),
\eeq
where $\chi_\alpha=\{{\rm p}\uparrow, \ {\rm p}\downarrow, \ {\rm n}\uparrow, \ 
{\rm n}\downarrow \}$ are the spin-isospin wave functions.
In the spatial part, the complex parameters
$ {\bf Z}_i ={\sqrt{\nu}({\bf D}_i+i{\bf K}_i/2\nu\hbar)}$
represent phase space coordinates.
In this work, the size parameter $\nu = 0.27$ ${\rm fm}^{-2}$ is chosen
to reproduce the energy and the r.m.s. radius of the $\alpha$ particle.

%
% Constraint Cooling  --- Generate GCM basis states before J-pi Proj.
%
In order to generate various intrinsic states,
we employ the constraint cooling
method~\cite{Horiuchi92,Enyo95,Enyo95b,Enyo95c},
by which we can search for a set of suitable complex parameters 
$\{ \bf Z_i \}$ to minimize the expectation value of the energy under some
constraints.
Following Ref.~\cite{Enyo95}, we obtain an optimum configuration state
by solving the constraint cooling equation given by
\begin{equation}
\label{CC}
{d{\bf Z}_i \over d{\beta}} 
	= 	-{\del \Hml \over \del {\bf Z}_i^\ast} 
  - \sum_{l=1} \eta_l {\del \langle W_l \rangle
\over \del {\bf Z}_i^\ast}\ ,
% {\rm and\ c.c.}\ ,
\end{equation}
where $\Hml$ is the expectation value of the Hamiltonian operator.
The expectation value of the constrained observable 
$\langle W_l \rangle$ in Eq.~(\ref{CC})
plays a role of the generator coordinate,
and $\eta_l$ is the Lagrange multiplier to keep 
the given constraint along the evolution.
In this work, we constrain the expectation value of the 
total oscillator quantum numbers $W_l=n_x$, $n_y$ and $n_z$ in a rectangular
coordinate system~\cite{Ita}.
It is possible to constrain other observables, such as an angular momentum,
however, cooling with an angular momentum constraint essentially results
in giving the rotating nucleus with the similar intrinsic state
when the angular momentum is not so high.
Then these states are not very effective as GCM basis states.

%
% J-pi Projection --- GCM basis states
%
Since the resultant cooled state is not the eigenstate of an angular momentum
in general,
it is necessary to carry out the angular momentum projection
to construct GCM basis wave functions in Eq.~(\ref{GCMwf}).
%This also applies to the linear-chain configuration in the next section.
\beq
\Phi^{JM} (\{ {\bf r}_i^{\alpha}; {\bf Z}_i \})
	= {2J+1 \over 8{\pi}^2 }
	{\int} d\Omega D^{J\ast}_{MK}(\Omega)
	\hat{R}(\Omega)\Phi (\{ {\bf r}_i^{\alpha}; {\bf Z}_i \}),
\eeq
where $\Omega$ is the Euler angle of $\{ {\bf r}_i^{\alpha} \}$.
In the numerical integration over the Euler angle
$\Omega(\alpha,\beta,\gamma)$,
we have adopted $25^3$ mesh points and integration is carried out with the 
Gauss-Legendre integral technique.
The intrinsic angular momentum component 
$K$ is chosen to be zero for simplicity.

%
%       Section 3
%
\section{$N\alpha$ linear-chain states in \nuc{12}{C} and \nuc{16}{O}}
\label{sec:results}

\subsection{3$\alpha$ states in \nuc{12}{C}} 

%
% Viewpoint of this work in 12C
%
In many previous analyses,
it has been shown that a large part of low-lying excited states of
\nuc{12}{C} have clustering structures.
Although Morinaga~\cite{Morinaga66} proposed 
the linear-chain structure for \nuc{12}{C},
the clustering states of $0_2^+$(7.66 MeV) and $2^+_2$($\sim$10 MeV)
have been discussed to have a loosely-coupled 
3$\alpha$ configuration~\cite{Uegaki80}
rather than the linear-chain structure.
We here investigate the stability of
the 3$\alpha$-linear-chain configuration in \nuc{12}{C}
from a viewpoint of orthogonality to low-lying states
by applying a microscopic 3$\alpha$ model.
In this calculation,
we use the Volkov No.2 potential~\cite{VolkovInt} with the Majorana value 
$M = 0.592$ as an effective nucleon-nucleon interaction. 
By using this interaction, the experimental binding energy 
of \nuc{12}{C} (7.4 MeV) from the 3$\alpha$ threshold,
which is calculated as $-83.69$ MeV, is well reproduced.

%
% GCM basis
%
Various intrinsic 3$\alpha$ configurations can be described
by a parameter set $(l_1,\ l_2,\ \theta)$,
where $l_1$ and $l_2$ are intervals between two $\alpha$-clusters and
$\theta$ is a bending angle (See Fig.1 a)).
The linear-chain configuration corresponds to $\theta=\pi$. 
Here,
we treat $l_1$ and $\theta$ as generator coordinates
and construct the basis states
which have various sets of these coordinates.
We employ $l_1 = 1,2,3,4,5$ fm and $\theta = \pi \sim \pi/3$. 
We regard $l_2$ as a variational parameter
and apply a cooling method to search for the energy minimum.
The 45 GCM basis states described by parameters ($l_1$, $l_2$, $\theta$) 
are obtained, and their explicit values are presented in Table I.

\FIGA

\TABLEA

%
% Pure Linear Chain
%
After projecting the obtained pure linear-chain intrinsic states ($\theta=\pi$)
to $J^{\pi}=0^+$,
we obtain the energy $-71.72$ MeV at the minimum point 
$(l_1,\ l_2)$ =(4 fm, 2.89 fm) (GCM basis No.4).
This energy is much higher than the 3$\alpha$ 
threshold where the $0^+_2$ state is observed.
%
% Longitudinal Vibration
%
To solve the longitudinal vibration of three $\alpha$ clusters,
we employ 5 GCM basis states which are given by
($l_1$, $l_2$) =
	(1 fm, 3.55 fm),
	(2 fm, 3.22 fm),
	(3 fm, 2.97 fm),
	(4 fm, 2.89 fm),
	and (5 fm, 2.92 fm),
keeping $\theta=\pi$ (GCM basis of No.1 $\sim$ No.5).
After solving the GCM equation,
we have the binding energy gain by 2.86 MeV ($-74.59$ MeV),
but its energy is still too short for the energy of the 
experimentary observed $0^+_2$ state.

%
% Bending Motion
%
Furthermore, we take account of the bending motion
	($\theta = 11\pi/12$, $10\pi$/12 and $9\pi/12$) 
which corresponds to fluctuation around the linear-chain configuration.
The solutions of the GCM equation 
with those basis states (No.1 $\sim$ No.20) give the lowest energy $-80.40$ MeV.
Although this energy is about 3 MeV higher
than the energy of the $0^+_2$ state,
it seems to give a good correspondence with that.
The wave function of this solution has a dominant components of
the linear-chain configuration,
which are seen from the fact that
overlap with the linear-chain GCM solution (described in the previous
paragraph) is 0.65.

%
% Orthogonality
%
From the analyses described above,
one may consider that the $0_2^+$ state can be interpreted as 
a 3$\alpha$ linear-chain state with small bending vibrations.
However, this solution may also have a large overlap
with the ground state of the $3\alpha$ system,
and it is necessary to solve the $0^+_2$ state by taking into account
of the coupling with the ground state.
% and a lower energy of the solution may be due to the mixing
% with the ground state.
%
%
%
%
In order to clarify this point, we performed the GCM calculation
with larger basis states including those which describe low-lying levels.
After diagonalizing the Hamiltonian with GCM basis states
(No.1 $\sim$ No.45 in table I),
we obtain the ground state energy $-91.23$ MeV which well corresponds
to the experimental one ($-92.2$ MeV).

\FIGB

%
% Stability
%
In Fig.2, we show the conversion behavior of the energy spectrum.
As the number of GCM basis states increases,
each basis configuration generally becomes closer
to equilateral triangle states.
The ground $0^+$ state is reproduced around the 
basis states No.30 which corresponds
to the configurations of $\theta = 6\pi/12$.
By this increase of basis states,
the second $0^+$ state converges energetically.
This solution has the overlap with linear-chain ($\theta=\pi$) basis states
of only 0.346 (absolute square),
%by only 0.346,
and cannot be interpreted as a linear-chain state.
%The result shows the $0^+$ state is not reproduced by a pure linear-chain
%but linear-chain with bending motion. <-I do not agree with this opinion.

\subsection{4$\alpha$ states in \nuc{16}{O}}

In a similar way to the 3$\alpha$ system,
we investigate the stability of the 4$\alpha$ linear-chain configuration
in \nuc{16}{O}.
We use the Volkov No.2 interaction~\cite{VolkovInt}
with the Majorana parameter $M=0.63$,
in order to reproduce 
the ground state energy of \nuc{16}{O} $(-127.6 {\rm MeV})$ ~\cite{Ita}.
However, this effective nuclear interaction has a shortcoming that the 
calculated binding energy of \nuc{12}{C} ($-82.0$ MeV) does not reproduce the 
observed one ($-92.2$ MeV).
This problem has been discussed~\cite{Ita} from both sides of model spaces
and effective interactions.
As one important conclusion,
it has been shown that the clustering states are well reproduced
around the corresponding threshold energy for various effective interactions.

\TABLEB

We apply the microscopic 4$\alpha$-cluster model and carry out 
the GCM calculation
for stability examination of the $4\alpha$ linear-chain configuration.
The employed GCM basis states are listed in Table II. 
The linear-chain states and their bending motions
are described by 45 GCM basis states 
(No.1 $\sim$ No.45 in Table II) of the
\nuc{8}{Be}+\nuc{8}{Be} configurations 
with length parameters ($l_1$ and $l_2$)
and an angle parameter ($\theta = \pi \sim \pi/2$)
as shown in Fig.1 (b). 
The relative position of two \nuc{8}{Be} clusters is 
determined by $l_2$ ($l_2 = 0.5, 1.0, 1.5, 2.0 $ and 2.5 fm)
and the distance $(l_1)$ between the two
$\alpha$-clusters of each \nuc{8}{Be} is chosen to minimize
the energy at given $l_2$ and $\theta$. 
In addition, we prepare 15 GCM basis states 
obtained by the Constraint Cooling Method
for the description of other lower states.

As the first step in research of the 4$\alpha$-linear-chain state,
we search the $0^+$ optimal linear-chain configuration with 
($l_1 = 3.29$ fm, $l_2 = 1.50$ fm, $\theta=0$) 
at the energy $E = 34.79$ MeV from the 4$\alpha$ threshold
within a single Slater determinant.
As the next step, we examine its stability for mixing with 
configurations other than the linear-chain by solving the GCM equation
with increasing the number of the basis states.
 
\FIGC

The $0^+$-state energy convergence is shown in Fig.3,
and the ground state energy converges to $-129.1$ MeV. 
The energy of the $0^+$ state obtained at $-91.68$ 
MeV (finally corresponds to $0^+_6$)
converges with a very small number of GCM basis states ($\cong$ No.15),
which means this state contains large linear-chain components
with small bending ($\leq 10\pi/12$).
As an evidence,
this state has an overlap (absolute square) 
with the GCM basis states Nos.1-10
	($\theta = \pi$ and $11\pi/12$)
	of 0.461
and
with the GCM basis states Nos.1-15
	($\theta = \pi, 11\pi/12$ and $10\pi/12$)
	of 0.613.
Even when a larger number of GCM basis states are adopted,
the state of $-91.68$ MeV is hardly affected by other lower states 
and approximately keeps its own energy value.
However this energy is higher about 20 MeV compared with the 
4$\alpha$ threshold energy ($-111.6$ MeV). 

There may be at least two reasons for the too-high-energy solution
of $0^+_6$ : 1) The \nuc{12}{C}+$\alpha$ threshold is not correctly 
reproduced due to
the present $NN$ interaction, and is calculated at the energy lower than
experiment by about 10 MeV. As was discussed in the previous paper,
it is very difficult problem to reproduce both threshold energies of
\nuc{12}{C}+$\alpha$ and 4$\alpha$ by an effective $NN$ interaction.
Because of this shortcoming, $4\alpha$ clusterized configurations are
calculated at rather high excited energies.
2) The GCM basis states describing the $0^+_6$ may be insufficient. We have
employed the configurations of bending motion deviating from the linear-chain,
and of the three-dimensional $4\alpha$ structure though they contribute
little to this $0^+_6$ state. Then the energy of this state would be
lower if we solve other small vibrations around the linear-chain configuration
such as $\cal{Z}$ letter-type mode or three-dimensional twinst mode.

\FIGD

However, an improved calculation based on the above reason for the fault
would only increase the quantative precision, and at least we can suggest
two things from the resent result.
First, the state of $-91.68$ MeV 
has a large overlap with the linear-chain configuration which shows the 
$\theta$ stability. Second, this state is hardly affected by other 
lower energy states and
there is a large possibility to correspond to remain as resonance
state. 

In Fig.4, the linear-chain states with different angular momenta are presented.
From the $2^+$ state, the rotational band is shown to appear
with a very small slope of 88 keV because of its large inertia moment.
This moment of inertia is similar to the experimentally suggested value
(about 60 keV).

\section{Summary and Conclusion}
\label{sec:summary}

In this peper,
we have microscopically studied the existence and the stability 
of $N\alpha$ linear-chain state in \nuc{12}{C} and \nuc{16}{O}.
We have emphasized the importance of 
the orthogonality between the linear-chain state and other lower states.
The linear-chain state of \nuc{12}{C} is hardly considered to be 
a stable state as we increase the number of the GCM basis states.
In other words, 
their components fragment to the levels over
broad range of excitation energy, especially to the $0_2^+$ state,
and these couplings make a pure linear-chain state unstable.
On the other hand,
the linear-chain state of \nuc{16}{O} is hardly affected by other states,
and the rotational band structure is shown to appear,
although the absolute energy of the band head is higher
by about 20 MeV compared with the 4$\alpha$-threshold.

These conclusions are consistent with the previous work
by Ikeda~\cite{Horiuchi72},
where the stability against small vibrations around
the equilibrium configuration is considered.
However, we have shown in this work that 
the linear-chain state of \nuc{16}{O} may survive
even after the couplings to the lower states are taken into account.

%%%%%%%%%%%%%%%%%%%%%%%%%%%%%%%%%%%%%%%%
% Acknowledgments
%%%%%%%%%%%%%%%%%%%%%%%%%%%%%%%%%%%%%%%%
\ack{
One of the authors (N.I.) thanks to 
the Japan Scientific Promotion Society for his fellowship.
The authors greatly thank to Professor K.Ikeda
for useful discussions, and to all the members of the Nuclear Theory Group
in Hokkaido University for great encouragements. 
}

%\vspace*{1cm}
%\centerline{\bf REFERENCES}

\newpage

\end{document}